# Double Blind Comparisons: A New Approach to the Database Aggregation Problem

**Abstract.** The Data Aggregation Problem occurs when a large collection of data takes on a higher security level than any of its individual component records. Traditional approaches of breaking up the data and restricting access on a "need to know" basis take away one of the great advantages of collecting the data in the first place.

This paper introduces a new cryptographic primitive, Double Blind Comparisons, which allows two co-operating users, who each have an encrypted secret, to determine the equality or inequality of those two secrets, even though neither user can discover any information about what the secret is.

This paper also introduces a new problem in bilinear groups, conjectured to be a hard problem. Assuming this conjecture, it is shown that neither user can discover any information about whether the secrets are equal, without the other user's co-operation.

We then look at how Double Blind Comparisons can be used to mitigate the Data Aggregation Problem. Finally, the paper concludes with some suggested possibilities for future research and some other potential uses for Double Blind Comparisons.

**Keywords:** Database aggregation, inference, private record linkage, privacy-preserving secure joins, Secure Multi-Party Computation

## 1 Database aggregation and inference

Data Aggregation and Inference have been recognized as a problem since at least 1982 [6], although standardized definitions have been lacking. NCSC Technical Report 005, Volume 1/5 [25] describes inference and aggregation as "different but related problems." It defines the inference problem as the problem "of users deducing (or inferring) higher level information based upon lower, visible data", following Morgenstern [23], and the aggregation problem as the problem that occurs "when classifying and protecting collections of data that have a higher security level than any of the elements that comprise the aggregate", following Denning et al. [10].

Both aggregation and inference can take place within a single database. This paper, however, deals with the situation where data is divided among two or more databases.

Denning et al. [10] identified two different types of aggregation. In this paper, we follow the terminology of White et al. [31] and refer to these as *cardinal aggregation* and *inference aggregation*. Although both types of aggregation can occur within a single database, we are concerned here with aggregation across multiple databases.

Cardinal aggregation occurs when an adversary collects a large number of similar records, each of which by itself is of little importance, but which by sheer volume become sensitive. An example of this often used in the literature (see e.g. [24]) is the CIA telephone directory, where each individual entry is of little significance, but the entire directory is considered classified.

Inference aggregation occurs when multiple databases are joined, creating virtual records with a large number of fields. While the sensitivity of the individual records may have been analyzed and found to be low, the sensitivity of the resulting virtual records has likely never been analyzed, and may be quite high.

As a simple example, consider the following two databases. The first is a military personnel database, containing two fields - a military Service Number (SN) and a job classification, with the SN as the primary key. The second database is a military medical database, containing the employee SN and a medical status, again with the SN as the primary key.

Consider the records for the member with SN C55-111-555. The personnel database shows this person to be a CF-18 Fighter Pilot. The medical database shows this person to be 4 months pregnant.

Neither of these facts, in isolation, provides any information about the operational effectiveness of the person's military unit. When combined, however, they allow an adversary to infer that there is a CF-18 fighter pilot who is currently unavailable to fly combat missions – information that might be of operational significance.

This research presents a new cryptographic primitive, Double Blind Comparisons, which allow two co-operating users, each in possession of a different ciphertext, to determine if the associated plaintexts are equal or not, even though neither user knows anything about the plaintext other than its encrypted value.

Neither user learns anything about the plaintext as a result of the comparison. With the exception of the trusted third party who issued the encrypted secrets, no one individual is able to carry out a comparison of two differing ciphertexts.

This primitive differs from Secure Multi-Party Computation [33]. With Secure Multi-Party Computation, the users each know their own secret and compute some function of the combined secrets, without revealing any information about their own secrets to the other parties; with Double Blind Comparisons, the users do not know the secrets they are comparing.

This is an important distinction, as it allows two users to determine whether or not their two secrets are equal. If the secret is known to one of the users, then the knowledge that the two secrets are equal immediately reveals the value of the other user's secret. This rules out Secure Multi-Party Computation as a solution.

Database inference may be made more difficult by increasing the difficulty of inference aggregation. For example, a government database may identify individuals by a Social Insurance Number, a bank by a customer ID number, and a business by an employee ID. Records from one database cannot then easily be linked to records from another. However, there are many cases where it is necessary to match a record from one database to its corresponding record in another.

Double Blind Comparisons will allow us to distribute a large dataset among multiple databases, while retaining the ability to match and compare records across two or more of these databases. For example, an employee's personnel records can be matched to their medical profile, but only if the administrators for the personnel and medical databases co-operate. Because the records cannot be linked otherwise, the security requirements on the two databases can be made much less stringent.

## 2 Preliminaries

### 2.1 Discrete Logarithm Problem

Given a group $G_1$ and two elements $g, h \in G_1$, the Discrete Logarithm Problem (DLP) is to find an element $x \in Z_q^*$ such that $h = g^x$ whenever such an element exists.

### 2.2 Diffie-Hellman Problem

The Computational Diffie-Hellman assumption states that, given $g, g^a, g^b \in G_1$, for a randomly-chosen generator $g$ and random exponents $a, b \in Z_p^*$, it is computationally intractable to compute the value $g^{ab}$. The associated problem of computing $g^{ab}$ is the Computational Diffie-Hellman Problem (CDHP).

The Decisional Diffie-Hellman assumption states that, given $g, g^a, g^b, g^c \in G_1$, for a randomly-chosen generator $g$ and random exponents $a, b \in Z_p^*$, it is computationally intractable to determine whether $g^c = g^{ab}$. The associated problem of determining whether $g^c = g^{ab}$ is the Decisional Diffie-Hellman Problem (DDHP).

## 2.3 Cryptographic Bilinear Map

A mapping $e(x,y): G_1 \times G_1 \to G_T$ is called a bilinear mapping if it satisfies the following properties:

Bilinear property: $e(x^a, y^b) = e(x,y)^{ab}$

Non-Degenerate. If $x, y$ are generators of $G_1$, then $e(x,y)$ is a generator of $G_T$.

Computability property: There is an efficient algorithm to compute $e(x,y)$ for all $x, y \in G_1$.

Our proof of security assumes the existence of a cryptographic bilinear map $e(x,y): G_1 \times G_1 \to G_T$ over groups $G_1, G_T$ of prime order p in which the Discrete Logarithm Problem is hard, and where the CDHP is hard in $G_1$.

## 2.4 Perfect Indistinguishability

We follow the terminology of Damgard and Nielsen [5].

Consider a probabilistic polynomial time (PPT) algorithm $U$. If we run $U$ on input string $x$, the output will be a probability distribution: for every possible string $y$ there some probability that $y$ is output when $x$ was the input. Call this probability $U_x(y)$. $U_x$ is the probability distribution of $U$'s output, on input $x$.

Next, consider two probabilistic algorithms $U, V$. We run both $U$ and $V$ on the same input $x$, and we choose one of the outputs produced, which we call $y$.

Definition: Given two PPT algorithms $U, V$, we say that $U, V$ are perfectly indistinguishable, written $U \sim^p V$, if $U_x = V_x$ for every $x$.

## 2.5 DDHP in bilinear groups

The DDHP is easily solved for group $G_1$ if a bilinear map $e(x,y)$ exists from $G_1 \times G_1 \to G_T$, as follows:

Given $g, g^a, g^b, g^c \in G_1$, it is easily shown that $g^c = g^{ab}$ iff $e(g^a, g^b) = e(g, g^c)$. It should be noted that this technique requires knowledge of the base g.

# 3 Related work

## 3.1 Secure Multi-Party Computation

Secure multi-party computation was initially suggested by A. C. Yao in a 1982 paper [33], in which he introduced the millionaire problem: Alice and Bob are two millionaires who want to find out who is richer without either revealing to the other the precise amount of their wealth. Yao proposed a solution allowing Alice and Bob to satisfy their curiosity while respecting the constraints. In general, Secure Multi-Party Computation allows $n$ users, each of whom has a secret $S_i$, to compute some function $F(\{S_i\}_{i=1}^n)$, without allowing user $U_i$ to learn the value $S_j$ for any $j \neq i$.

Because each user $U_i$ knows his or her own secret $S_i$, one problem that cannot be solved with secure multi-party computation is whether $S_i = S_j$ for any $j \neq i$, as this would immediately the value of $S_j$ to user $U_i$.

With Double Blind Comparisons, neither of the users knows the value of the secret they are comparing, and thus the question of equality can be resolved without compromising the secret.

## 3.2 Prior work on Database Aggregation and Inference

Data Aggregation and Inference were identified as problems at least as early as 1982 [5]. Ten years later, NCSC-TG-010 [24], included two short sections on the combined problem.

Early research focused mainly on Multi-Level Secure (MLS) database systems, in which both sensitive and non-sensitive data are stored; the MLS system relies on classification rules and access controls to ensure that sensitive data is released only to an appropriately cleared user. The problem then was to ensure that the results of multiple unclassified queries would not allow an uncleared user to infer classified data. NCSC TECHNICAL REPORT – 005 Volume 1 of 5 [25] dealt with this in greater detail.

This early work focused entirely on preventing the inference of sensitive data from non-sensitive queries. It was tacitly assumed that the sensitive data would be available from other sources. However, as database technology has improved dramatically over the past two decades, the advantages of large scale unclassified databases have started to override the security concerns of allowing smaller databases to be combined.

For example, and especially since 9/11, multi-agency security teams are commonly established for any large scale event which is a credible terrorist target. Each agency may have a file on a particular individual, but for security/privacy reasons, they are unable to simply join their databases. How, then, can they determine whether this individual constitutes a threat, and if so, how serious a threat?

## 3.3 Prior work based on Privacy Preservation

A closely related problem from the field of privacy is the study of privacy-preserving databases. In this case, the objective is to allow the data to be used (e.g. for statistical or research purposes) while preventing the release of personally identifying information.

**Quasi-identifiers, $k$-anonymity, $l$-diversity, and $t$-closeness**

Sweeney [30] showed the existence of *quasi-identifiers* – groups of attributes that had the potential to uniquely identify an individual, even though the data had supposedly been anonymized. [30] demonstrated that approximately 87% of the US population can be uniquely identified by the combination of three attributes – date of birth, 5-digit zip code, and gender. In [29], Sweeney was able to re-identify the medical records for then-Governor of Massachusetts William Weld, based on these three attributes.

To address this, the techniques of $k$-anonymity [26][27][28][29], $l$-diversity [18], and $t$-closeness [14] were developed. These techniques are focused on preventing the joining of multiple tables; in effect, they address the inference aggregation problem by preventing the join from being made. As such, these techniques are separate from, but complementary to, our proposed technique, which is designed to allow secure matching of corresponding records to be made in controlled circumstances.

**Information Sharing Across Private Databases**

Agrawal et al. [1] developed techniques to enable two co-operating users to create secure joins and intersections of database tables, using commutative encryption. However, their method relies on the databases themselves being adequately protected, whereas our technique requires only that the users' private keys remain protected.

**Private Record Linkage**

Record Linkage is the problem of matching corresponding records from disparate collections, where there may be no unique identifiers, using statistical methods. Record linkage typically assumes there are no keys that uniquely identify individuals across distinct datasets; instead, linkage relies on probabilistic methods and machine learning algorithms.

For example, a personnel database gives the following information for Captain Juanita Mendes Smith, CF-18 pilot, age 32, auburn hair, hazel eyes, height 5' 2", weight 118 pounds. Is this the same

Mrs. J. M. Smith, early 30's, brown hair, green eyes, height 5' 2", weight 140 pounds, 4 months pregnant, listed in the local hospital database?

Private record linkage [32] is the problem of matching such corresponding records without releasing any personally identifiable information. Our research, by comparison, focuses on the matching of records for which encrypted unique identifiers exist.

**Privacy-Preserving Joins**

Kantarcioglu et al. [13] addresses the issue of matching pairs of records which apply to the same subject, while precluding invasion of that subject's privacy. In this work, all records are encrypted by the researcher who gathers the data, using the public key of a Trusted Third Party (TTP) known as the Key Server (KS). All records are stored in their encrypted form on a central Data Storage (DS) database.

The Data Storage administrator can determine to a high degree of certainty when two records refer to the same subject, but cannot decrypt the records. After identifying candidate pairs of records, DS submits them to the TTP who is the only one who can decrypt the records.

The TTP will return the requested information only if the records are, indeed, referring to the same subject.

Disadvantages of this system include the reliance on the separation of duties and the existence of a trusted third party, and the concern that the DS and the KS may collaborate.

If KS could gain access to the entire DS database, it would be a simple matter to decrypt all the data. Thus, an enormous amount of trust would have to be placed in KS, simply to protect what is essentially unclassified data. So, as with Agrawal et al. [1] this method requires the database to be strongly protected, whereas our technique requires protection only for the users' private keys.

## 4  Proposed Solution

### 4.1  Conjecture

Let $G_1, G_T$ be cyclic groups of prime order p in which the Discrete Logarithm Problem is hard, for which $e(x, y): G_1 \times G_1 \to G_T$ is a cryptographic bilinear map.

The following problem is conjectured to be hard:

Given $h, h^N, g^{rs}, g^{rsN}, g^x \in G_1$, where $G_1 = <g> = <h>$, determine whether $g^{sN} = g^x$.

This can be rewritten as

Given $h, h^N, g_s^r, g_s^{rN}, g_s^{x'} \in G_1$, where $g_s = g^s, G_1 = <g_s> = <h>, x' = x/s$, determine whether $g_s^N = g_s^{x'}$.

Let O be an oracle that, on input $\{h, h^N, g_s^r, g_s^{rN}, g_s^{x'}\}$, can determine whether $g_s^N = g_s^{x'}$. This is equivalent to determining whether $g_s^{rN} = g_s^{rx'}$, which is the DDHP, but without knowledge of the base $g_s$.

It is known that the DDHP is easy in bilinear groups if the base is known, but the technique used does not work if the base is unknown.

Therefore, I conjecture, but was unable to prove, that the above problem is hard.

### 4.2  Overview

In discussing this system, we consider the following parties: a submitter (Alice), a responder (Bob), and a Trusted Central Authority (Ted).

Alice and Bob each maintain a database of items; each item $i$ is assigned a unique identifier $N_i$. However, none of them know the value $N_i$; in their databases, the item is identified by a one-way encryption of $N_i$; respectively $E_a(N_i)$, $E_b(N_i)$.

In his database, Ted maintains the link between the item i and its identifier $N_i$. e.g. {i = Captain Juanita Mendes Smith; $N_i$ = C55-111-555}. Ted is the only one who knows the values {$N_i$}.

In her (resp. his) database, Alice (resp. Bob) uses the value $\alpha_j = E_a(N_i)$ (respectively $\beta_k = E_b(N_i)$ ) as the index and stores data about item i (e.g. military occupation qualifications, or medical status) but not the item identifier. E.g. {index = $\alpha_j$ = $E_a(N_i)$; Military occupation = CF-18 pilot}.

Note that the indices j, k of $\alpha_j$, $\beta_k$ bear no association with the index i of $N_i$, or with each other.

### 4.3 Set-up

Let $G_1, G_T$ be cyclic groups of prime order $p$ in which the Discrete Logarithm Problem is hard, for which $e(x, y): G_1 \times G_1 \to G_T$ is a cryptographic bilinear map.

Let $M, N$ be integers known only to Ted.

Let $(g_a, s)$ (respectively $(g_b, t)$) be Alice's (respectively Bob's) private key. ($g_a, g_b$ are generators of $G_1$.)

For a given identifier M:

Alice chooses $r \in_R Z_p^*$ and sends $g_a^r$ to Ted.

Ted computes and sends $g_a^{rM}$ to Alice.

Alice chooses $i$, where the $i^{th}$ record is the next available free entry in database $DB_a$. She computes $s_i = E_s(i)$ for some suitable encryption function E with key $s$ (e.g. AES). She also computes $\alpha_i = (g_a^{rM})^{s_i/r} = g_a^{s_i M}$ and stores it as the index attribute for the $i^{th}$ record in database $DB_a$. Using $s_i$ rather than $s$ ensures that each record index is encrypted with a different exponent.

Bob likewise communicates with Ted and stores $\beta_j = (g_b^{rM})^{t_j/r} = g_b^{t_j M}$ as the index attribute for the $j^{th}$ record in database $DB_b$, where the $j^{th}$ record is the next available free entry in $DB_b$.

### 4.4 Double Blind Comparison

Define $\alpha_i \approx \beta_j$ if $s^{-1} \log_{g_a}(\alpha_i) = t^{-1} \log_{g_b}(\beta_j)$ ; i.e. $\alpha_i = g_a^{s_i M}$ and $\beta_j = g_b^{t_j M}$ are both one-way encryptions of the same (hidden) value M .

To determine whether $\alpha_i \approx \beta_j$ , where $\alpha_i = g_a^{s_i N_{\sigma(i)}}$ and $\beta_j = g_b^{t_j N_{\varphi(j)}}$ , Alice chooses a random number $r$ and sends Bob $(g_a^{rs_i}, \alpha_i^r)$ .

Bob calculates:
$$B_0 = e(g_a^{rs_i}, \beta_j) = e(g_a, g_b)^{rs_i t_j N_{\varphi(j)}}$$
$$B_1 = e(\alpha_i^r, g_b^{t_j}) = e(g_a, g_b)^{rs_i t_j N_{\sigma(i)}}$$

$\alpha_i \approx \beta_j$ if and only if $N_{\varphi(j)} = N_{\sigma(i)}$ if and only if $B_0 = B_1$

### 4.5 Example

To give an example of how this works, suppose Carol is a flight dispatcher who is trying to put together a low-level CF-18 reconnaissance mission. She needs to know if Captain Juanita "Joan"

Smith is qualified and capable to fly CF-18s on such a mission. Alice is the database operator in charge of the base personnel database, and Bob is in charge of the base hospital medical records.

In her database, Carol uses the value $\gamma_l = E_c(N_l)$ as the index and stores the item identifier. E.g. {index = $\gamma_l = E_c(N_l)$; $l$ = Juanita Smith; $N_l$ = C55-111-555}.

In this example, the protocol is invoked twice – once with Carol as the submitter and Alice as the responder, and a second time with Alice as the submitter and Bob as the responder.

First invocation – Carol to Alice

1. Carol looks up Captain Smith's identifier $\gamma_i$, chooses a random exponent $r$, and sends $g_c^{ru_i}, \gamma_i^r$ where $(g_c, u)$ is Carol's private key, to Alice, along with a request to know if this individual is currently qualified to fly CF-18s on combat missions.
2. Alice searches through her database until she finds $\alpha_j \approx \gamma_i$. (This is extremely inefficient for large databases. Later we will look at ways to modify this technique to conduct binary searches.) She confirms that $\alpha_j$ is qualified and current as a pilot on CF-18s. She next needs to verify that $\alpha_j$ is medically fit to fly combat missions.

Second invocation – Alice to Bob

Alice chooses a random exponent $r'$ and sends $g_a^{r's_j}, \alpha_j^{r'}$ to Bob, along with a request to know if this individual is medically fit to fly on combat missions.

Bob searches through his database until he finds $\beta_k \approx \alpha_j$. He finds that $\beta_k$ is 4 months pregnant.

Response to second invocation – Bob to Alice

Bob advises Alice that employee $\alpha_j^{r'}$ is limited to restricted flight duties (commercial and pressurized transport aircraft only; no aerobatics).

Response to first invocation – Alice to Carol

Alice advises Carol that employee $\gamma_i^r$ is not able to fly CF-18 combat missions.

Alternatively, Carol could contact Bob directly to obtain Captain Smith's medical flight status.

Points to note:

- Carol does not know the reason why Captain Smith is not available to fly the mission. In particular, she does not find out that her friend Juanita is pregnant, as she has no need to know this information.
- Alice does not know the identity of the individual for whom the query was issued. She learns nothing beyond the fact that this individual is currently not medically cleared to fly combat missions.
- Bob learns nothing about the individual beyond what is already contained in his own database, although - depending on the wording of the query - he may be able to infer that this individual is a flight crew member (e.g. pilot, navigator, or flight attendant).

## 5  Security

First of all, note that given $\alpha_i = g_a^{s_i N_{\sigma(i)}}$, even if $s_i, g_a$ are known, recovering $N_{\sigma(i)}$ is equivalent to solving the DLP. So other than Ted, no user of the system knows the values $\{N_i\}$.

Ted is never given the values of $g_a$ or $g_a^{s_i}$.

We demonstrate the correctness. Informally, correctness means that an honest submitter, submitting a request to an honest responder, will receive a correct result if and only if the request is valid.

We then analyze the following attacks:

- Alice knows the contents of DB$_a$ and her own secret key $(g_a, s)$; she obtains a copy of Bob's database DB$_b$. Alice succeeds if, without Bob's co-operation, she can choose any $\alpha_i \in$ DB$_a$ and confirm which $\beta_j \in$ DB$_b$ corresponds to $\alpha_i$.
- Bob knows the contents of DB$_b$ and his own secret key $(g_b t)$; he obtains a copy of Alice's database DB$_a$. He receives a query $(g_a^{rs}, \alpha^r)$ from Alice, from which he is able to determine $\beta_j \in$ DB$_b$ such that $\beta_j \approx \alpha$. Bob succeeds if he can determine which $\alpha_i \in$ DB$_a$ is the subject of Alice's query.
- Ted knows all the $\{N_i\}$, and obtains a copy of Alice's database DB$_a$. Ted succeeds if, given any $N_I$ and any $\alpha_J \in$ DB$_a$, he can determine whether $\alpha_J = g_a^{s_J N_I}$.

## 5.1 Correctness

Let $\alpha$ be an entry in DB$_a$, and $\beta$ be an entry in DB$_b$, where $\alpha = g_a^{sN_i}$ and $\beta = g_b^{tN_i}$; i.e. $\alpha \approx \beta$. Alice chooses $r \in_R L$ and sends $(g_a^{rs}, \alpha^r)$ to Bob. (For convenience, we leave out the subscripts for $\alpha, s, \beta$ and $t$.)

Then Bob can demonstrate, without knowledge of Alice's secret key $(g_a, s)$ but with knowledge of his own secret key $(g_b, t)$, that $\alpha \approx \beta$.

Conversely, if $e(\alpha^r, g_b^t) = e(g_a^{rs}, \beta)$, where $\alpha = g_a^{sM}$ and $\beta = g_b^{tN}$, then $M = N$.

*Proof:*

$$\alpha = g_a^{sN_i} \text{ and } = g_b^{tN_i}.$$

$$e(\alpha^r, g_b^t) = e(g_a^{sN_i r}, g_b^t) = e(g_a, g_b)^{rstN_i} \text{ and}$$

$$e(g_a^{rs}, \beta) = e(g_a^{rs}, g_b^{tN_i}) = e(g_a, g_b)^{rstN_i}$$

Conversely, suppose $e(\alpha^r, g_b^t) = e(g_a^{rs}, \beta)$, where $\alpha = g_a^{sM}$ and $\beta = g_b^{tN}$. Then $M = N$ as follows:

$$e(g_a^{rs}, \beta) = e(g_a^{rs}, g_b^{tN}) = e(g_a, g_b)^{rstN} \text{ and}$$

$$e(\alpha^r, g_b^t) = e(g_a^{sMr}, g_b^t) = e(g_a, g_b)^{rstM}$$

Since $g_a^{rs}, g_b^t$ are generators of $G_1$, $e(g_a^{rs}, g_b^t) = e(g_a, g_b)^{rst}$ is a generator of $G_T$, and so $M = N$.

## 5.2 Unlinkability

To prove unlinkability, we prove that neither Alice nor Bob, given full disclosure of both databases DB$_a$ and DB$_b$, can obtain an advantage in linking any pair of records between the two databases without the other's co-operation. In this, Alice is the submitter, while Bob is the responder in a transaction initiated by Alice.

**Alice**

Assume Alice knows her own secret key, plus all the entries $\alpha_i$ in DB$_a$, and all the entries $\beta_j$ in DB$_b$.

Let U, V be PPT algorithms as follows:

U, given input $\alpha = \alpha_i$ for some specific $\alpha_i$ in DB$_a$, returns the unique entry $\beta_U$ in DB$_b$ for which $\approx \beta_U$.

V, given input $\alpha = \alpha_i$ for some specific $\alpha_i$ in $DB_a$, returns an entry $\beta_V$ in $DB_b$ which does not correspond to $\alpha$ in $DB_a$.

Then $U \sim^p V$.

*Proof:*

For any set $g_a, g_a^s, g_b$, $\alpha = g_a^{sM_U}$, $\beta_U = g_b^{tM_U}$, and $\beta_V = g_b^{tM_V}$, with $M_U \neq M_V$, show that there exist $u, N \neq M_U$ such that $\beta_U = g_b^{uN}$, $\beta_V = g_b^{uM_U}$.

Because $g_b^{M_U}$ is a generator, $\exists u$ such that $(g_b^{M_U})^u = \beta_V$; then $u = tM_V M_U^{-1}$ exists with $M_U = tM_V$. Since $M_U \neq M_V$, $u \neq t$.

Set $= u^{-1} tM_U$; then $uN = tM_U$; then $\alpha = g_a^{sM_U}$, $\beta_U = g_b^{uN}$, and $\beta_V = g_b^{uM_U}$, with $M_U \neq N$.

Let O be an oracle that, on input $\{g_a, g_a^s, \alpha, \beta_U, \beta_V\}$, with either $\alpha \approx \beta_U$ or $\alpha \approx \beta_V$, outputs $W \in \{U, V\}$.

Let $PR_O = \text{pr}\{g_a, g_a^s, \alpha, \beta_U, \beta_V, : \alpha \approx \beta_W\}$; in other words, $PR_O$ is the probability that O outputs the correct answer. When $\alpha \approx \beta_U$, a correct answer will be $W = U$, and an incorrect answer will be $W = V$. When $\alpha \approx \beta_V$, a correct answer will be $W = V$ and an incorrect answer will be $W = U$.

Let $PR_{O(g_b, t, M_U, M_V)}$ be the probability that O outputs the correct answer when Bob's secret key is $(t, g_b)$, $\beta_U = g_b^{uM_U}$, and $\beta_V = g_b^{tM_V}$.

Let $PR_{O(g_b, u, N, M_U)}$ be the probability that O outputs the correct answer when Bob's secret key is $(u, g_b)$, $\beta_U = g_b^{uN}$, and $\beta_V = g_b^{uM_U}$.

Since the output of O is independent of $(g_b, t, M_U, M_V)$ and $(g_b, u, N, M_U)$, it follows that $PR_{O(g_b, t, M_U, M_V)} = PR_{O(g_b, u, N, M_U)} = PR_O$.

However, a correct answer when the secret values are $(t, M_U, M_V)$ is an incorrect answer when those values are $(u, N, M_U)$.

Therefore, $PR_{O(g_b, t, M_U, M_V)} = 1 - PR_{O(g_b, u, N, M_U)}$; i.e. $= PR_O = 1 - PR_O$, or $PR_O = 0.5$.

Therefore, Alice has no advantage in distinguishing $U$ from ; i.e. $U \sim^p V$.

∎

## Bob

Bob receives a request $(g_a^{rs_x}, \alpha_x^r)$ from Alice, where $x$ is unknown to him.

Assume Bob knows all the entries $\alpha_i$ in Alice's database $DB_a$, in addition to his own secret key and all the entries $\beta_j$ in $DB_b$. Bob can determine that Alice is inquiring about the entity $\beta_J$ for some particular $\beta_J$. He wants to know which record $\alpha_i$ is the subject of Alice's inquiry.

In particular, if Bob picks a specific record $\alpha_K \in DB_a$, he succeeds if he can determine whether $\alpha_K \approx \beta_J$.

Assuming conjecture 4.1, this is a hard problem.

*Proof:*

$\beta_J = g_b^{t_J N \varphi(J)}$. Since Bob knows $t_J$, he can easily compute $g_b^{N \varphi(J)}$. Also, since Bob knows that $\alpha_x \approx \beta_J$, he knows that $\alpha_x^r = g_a^{s_x N \varphi(J)}$.

Thus Bob knows $g_b, g_b^{N_{\varphi(J)}}, g_a^{rs}, g_a^{rsN_{\varphi(J)}}, \alpha_K = g_a^{s_K N_{\sigma(K)}}$, and wishes to determine whether $g_a^{sN_{\varphi(J)}} = g_a^{s_K N_{\sigma(K)}}$.

By conjecture 4.1, this is a hard problem.

∎

**Ted**

Ted knows all the $\{N_i\}$ and all the $\{\alpha_j\}$ from $DB_a$. Ted succeeds if, given any pair $N_I, \alpha_J$, he can determine whether $\alpha_J = g_a^{s_J N_I}$. We present an outline of a proof that Ted is unable to do so.

Let U, V be PPT algorithms as follows:

U, given input $N_I$, returns the unique entry $\alpha_U$ in $DB_a$ for which $\alpha_U = g_a^{s_U N_I}$.

V, given input $N_I$, returns an entry $\alpha_V$ in $DB_a$ for which $\alpha_V \neq g_a^{s_U N_I}$.

For any set $\{N_I, \alpha_U = g_a^{s_U N_I}, \alpha_V = g_a^{s_V N_J}\}$, with $N_I \neq N_J$, there exist $u, v$ such that

$$\alpha_U = g_a^{u N_J}, \alpha_V = g_a^{v N_I}$$

Simply set $u = s_U N_I / N_J$ and $v = s_V N_J / N_I$.

Because $s_U = E_s(U), s_V = E_s(V)$ are symmetric encryptions of $U, V$ respectively, the probability distributions of $u, s_U, v, s_V$ are computationally indistinguishable.

Therefore, Ted is computationally unable to distinguish between U and V.

A more detailed proof will be forthcoming.

# 6 Future work

## 6.1 Efficiency

As mentioned, the simplistic approach given is very inefficient. A proposal to improve the efficiency would be to identify each record by several keys, which would allow a binary search.

For example, take a database with 8 records - $\theta_{000}$ through $\theta_{111}$.

Alice's secret key would be $(s_1, g_a^{s_1}, s_2, g_a^{s_2}, s_3, g_a^{s_3})$.

Set $\alpha_{0,1} = g_a^{sN_1}, \alpha_{0,2} = g_a^{sN_2}, \alpha_{0,3} = g_a^{sN_3}, \alpha_{1,1} = g_a^{sM_1}, \alpha_{1,2} = g_a^{sM_2}, \alpha_{1,3} = g_a^{sM_3}$.

Then record $\theta_{010}$, for example, would be identified by $\alpha_{0,1}, \alpha_{1,2}, \alpha_{0,3}$ ; while record $\theta_{110}$ would be identified by $\alpha_{1,1}, \alpha_{1,2}, \alpha_{0,3}$.

Alice's secret key would be $(s_1, g_a^{s_1}, s_2, g_a^{s_2}, s_3, g_a^{s_3})$.

To inquire about record $\theta_{101}$ , Alice chooses $(r_1, r_2, r_3) \in_R L$ and sends $(g_a^{s_1 r_1}, \alpha_{1,1}^{r_1})$, $(g_a^{s_2 r_2}, \alpha_{0,2}^{r_2})$, and $(g_a^{s_3 r_3}, \alpha_{1,3}^{r_3})$ to Bob.

A more detailed analysis of this improvement is out of scope and will be dealt with in a future paper.

## 6.2 Cardinal Aggregation

Cardinal aggregation can also be dealt with by dividing the information up on a "need to know" basis among several different databases. For example, the personnel database for the entire Canadian Armed Forces might be split up among the different military bases and ships. The Personnel Officer for CFB Cold Lake would have no access to the personnel database for HMCS Halifax.

Double Blind Comparisons might be useful in this scenario for synchronizing databases when a service member is being posted from one unit to another. Further analysis of this scenario will be dealt with in a future paper.

## 6.3 Anonymous Credentials

An electronic credential is issued to a user by one organization (the issuer) that enables the user to demonstrate to a third party (the verifier) that the user possesses some attribute. With anonymous credentials [2][3][4], a user is able to obtain a credential from an issuing organization (possibly using a pseudonym) and use that credential to prove possession of some attribute to multiple verifiers, using different pseudonyms, in such a way that the issuer of the credential and the verifiers would be unable to link the transactions.

Double Blind Comparisons might enable one organization (Alice) to prove to another organization (Bob) that a user (Carol) possesses some attribute or attributes (e.g. a Secret security clearance and a cryptographic public key) without revealing any more information to either Alice or Bob than is strictly necessary.

Further analysis of this usage will be dealt with in a future paper.